\documentclass[aps,showpacs,prd,twocolumn]{revtex4}
\usepackage{graphicx}
\usepackage{amssymb}
\usepackage{amsmath}
\usepackage{epstopdf}
\usepackage{color}
\begin{document}

\def\SU2{\ensuremath{{SU(2)}}}
\def\sdmass{\ensuremath{\sqrt{a_o}/2}}
\newcommand{\bea}{\begin{eqnarray}}
\newcommand{\eea}{\end{eqnarray}}
\def\half{\frac{1}{2}}
\def\beq{\begin{equation}}
\def\eeq{\end{equation}}
\def\beqn{\begin{eqnarray}}
\def\eeqn{\end{eqnarray}}

 \title{Stability of self-dual  black holes}
\author{Eric Brown, Robert Mann}
 \affiliation{Department of Physics and Astronomy, University of Waterloo, Waterloo, Ontario N2L 3G1, Canada}
\author{Leonardo Modesto}
\affiliation{Perimeter Institute for Theoretical Physics, 31 Caroline St.N., Waterloo, ON N2L 2Y5, Canada }


\begin{abstract}

We study the stability properties of the Cauchy horizon for 
two different self-dual
black hole solutions obtained in a model inspired by Loop Quantum Gravity.
The self-dual spacetimes depend
on a free dimensionless parameter called a {\em polymeric} parameter $P$.
For the first metric the Cauchy horizon is stable for supermassive
black holes only if this parameter is sufficiently small. For small black holes, however the stability
is easily implemented.
The second metric analyzed is not only self-dual but also {\em form-invariant} under the
transformation $r \rightarrow r_*^2/r$ and $r_* = 2 m P$. We find that this symmetry protects the Cauchy horizon
for any value of the polymeric parameter.


\end{abstract}

\pacs{04.60.Bc:, 04.70.Dy, 98.70.Sa}

\maketitle
\tableofcontents
\section{Introduction}

One approach to quantum gravity, Loop Quantum Gravity
(LQG) \cite{LQGgeneral,LQGgeneral2,LQGgeneral3}, has given rise to
models that afford a description of  the very early universe. This simplified
framework, which uses a minisuperspace approximation, has been shown
to resolve the initial singularity problem \cite{Bojowald}.
 A black hole metric in this model, known as the loop black hole (LBH)
 \cite{Modesto:2008im}, has  a property of self-duality
that removes the singularity and replaces it with
another asymptotically flat region. Both the
thermodynamic properties  \cite{Modesto:2008im,poly} and the dynamical aspects of  collapse and
evaporation \cite{Hossenfelder:2009fc} of these self-dual black holes have been previously studied.
These black hole spacetimes have also been investigated in a midi-superspace reduction of LQG \cite{GP}.

The LBH has two horizons -- an event horizon and a Cauchy horizon -- and as such raises additional questions as to the stability of its interior \cite{RNstability,poisson}.  Cauchy horizons are notoriously unstable, and it is not a-priori clear that the LBH has a stable interior.
In the present work we consider this question by analyzing the behaviour of
a scalar field propagating inside the outer horizon.  We find that the LBH has improved
stability over classical 2-horizon black holes, such as the Reissner-Nordstr\(\ddot{\text{o}}\)m black hole.
We find furthermore that {\em a particular subclass of} LBHs {\em are fully stable} under such perturbations.

Our paper is organized as follows. First, we
recall the loop black hole (LBH) derivation in short.
Second, we derive the equation of motion for a scalar field in the LBH
background then we derive the solution near the horizons.
We conclude by applying the analysis to two different kind of LBHs showing
a substantial improvement of the stability of the Cauchy horizon.
The two metrics depend on a free parameter $P$ known as the {\em polymeric parameter}.
For the first LBH metric we obtain stability if $P$ is sufficiently small.  
The second metric instead is stable for any value of $P$.

\section{Loop black hole}
\label{metric}

The regular black hole metric that we will be using is derived from a simplified model of LQG \cite{Modesto:2008im}.
LQG is based on a canonical quantization of the Einstein equations written in terms of the Ashtekar variables \cite{AA}, that is in terms of an $SU(2)$ 3-dimensional connection $A$ and a triad $E$. The basis states of LQG then are closed graphs, the edges of which are labeled by irreducible $SU(2)$ representations and the vertices by $SU(2)$ intertwiners (for a review see e.g. \cite{LQGgeneral,LQGgeneral2,LQGgeneral3}). The
edges of the graph represent quanta of area with area $\gamma l_P^2 \sqrt{j(j+1)}$, where $j$ is a half-integer representation label on the edge, $l_P$ is the Planck length,  and $\gamma$ is a parameter of order $1$ called the Immirzi parameter. The vertices of the graph represent quanta of $3$-volume. One important consequence that we will use in the following is that the area is quantized and the smallest possible quanta correspond to an area of $\sqrt{3}/2 \gamma l_P^2$.

To obtain the simplified black hole model the following assumptions were made. First, the number of variables was reduced by assuming spherical symmetry. Then, instead of all possible closed graphs, a regular lattice with edge lengths $\delta_b$ and $\delta_c$ was used. The solution was then obtained dynamically inside the homogeneous region (that is inside the horizon where space is homogeneous but not static). An analytic continuation to the outside of the horizon shows that one can reduce the two free parameters by identifying the minimum area present in the solution with the minimum area of LQG.
The one remaining unknown constant $\delta_b$  is a
dimensionless parameter of the model that determines the strength of deviations from the classical theory, and would have to be constrained by experiment. Redefining  $\delta_b = \delta$, the free parameter that appears in the metric is 
$\epsilon = \delta\gamma$, the product of the Immirzi parameter $\gamma$  and the polymeric quantity $\delta$.
With the plausible expectation that   quantum gravitational corrections become relevant only when the curvature is in the Planckian regime, corresponding to $\delta \gamma  < 1$, outside the horizon the solution is the Schwarzschild solution up to negligible Planck-scale corrections in $l_P$ and $\delta \gamma$.
This quantum gravitationally corrected
Schwarzschild metric can be expressed in the form
\begin{eqnarray}
ds^2 = - G(r) dt^2 + \frac{dr^2}{F(r)} + H(r) d\Omega^{(2)},
\label{g}
\end{eqnarray}
with $d \Omega^{(2)} = d \theta^2 + \sin^2 \theta d \phi^2$ and
\begin{eqnarray}
&& G(r) = \frac{(r-r_+)(r-r_-)(r+ r_{*})^2}{r^4 +a_o^2}~ , \nonumber \\
&& F(r) = \frac{(r-r_+)(r-r_-) r^4}{(r+ r_{*})^2 (r^4 +a_o^2)} ~, \nonumber \\
&& H(r) = r^2 + \frac{a_o^2}{r^2}~ .
\label{statgmunu}
\end{eqnarray}
Here, $r_+ = 2m$ and $r_-= 2 m P^2$ are the two horizons, and $r_* = \sqrt{r_+ r_-} = 2mP$. $P$ is the
polymeric parameter $P = (\sqrt{1+\epsilon^2} -1)/(\sqrt{1+\epsilon^2} +1)$, with
$\epsilon = \delta\gamma \ll 1$. Hence $P \ll 1$,  implying $r_-$ and $r_*$ are very close to $r=0$. The area $a_o$ is equal to $A_{\rm min}/8 \pi$, $A_{\rm min}$ being the minimum area gap of LQG.

Note that in the above metric, $r$ is only asymptotically the usual radial
coordinate since $g_{\theta \theta}$ is not just $r^2$.  We shall see that this choice of
coordinates however has the advantage of easily revealing the properties
of this metric. Most importantly, in the limit
$r \to \infty$, the deviations from the Schwarzschild-solution are of
order $M \epsilon^2/r$, where $M$ is the usual ADM-mass:
\beqn
G(r) &\to& 1-\frac{2 M}{r} (1 - \epsilon^2)~, \nonumber  \\
F(r) &\to& 1-\frac{2 M}{r}~ , \nonumber \\
H(r) &\to& r^2 .
\eeqn
The ADM mass is the mass inferred by an observer at flat asymptotic infinity; it is determined solely
by the metric at asymptotic infinity.  The parameter $m$ in the solution is related to the mass $M$ by $M = m (1+P)^2$.

If one now makes the coordinate transformation $R = a_o/r$ with the rescaling
$\tilde t= t \, r_*^{2}/a_o$, and
simultaneously substitutes $R_\pm = a_o/r_\mp$, $R_* = a_o/r_*$ one finds that the metric in
the new coordinates has the same form as in the old coordinates, thus exhibiting a
very compelling type of self-duality with dual radius $r=\sqrt{a_o}$. Looking at the angular part
of the metric, one sees that this dual radius corresponds to a minimal possible
surface element. It is then also clear that in the limit $r\to 0$, corresponding
to $R\to \infty$, the solution
does not have a singularity, but instead has another asymptotically flat Schwarzschild region.

An important quantity for studying stability is the surface gravity
\begin{eqnarray}
\kappa^2 = - g^{\mu \nu} g_{\rho \sigma} \nabla_{\mu} \chi^{\rho} \nabla_{\nu}
\chi^{\sigma}/2 = - g^{\mu \nu} g_{\rho \sigma}  \Gamma^{\rho}_{\;\mu 0} \Gamma^{\sigma}_{\;\nu 0}/2,
\end{eqnarray}
where $\chi^{\mu}=(1,0,0,0)$ is a timelike Killing vector 
in $r>r_+$ and $r<r_-$ but space-like 
in $r_- <r < r_+$ and $\Gamma^{\mu}_{\; \nu \rho}$
are the connection coefficients.  
%
%
For the metric (\ref{g}) we  find the following values 
\begin{eqnarray}
 \kappa_- =  \frac{4 m^3 P^4 (1-P^2)}{16 m^4 P^8 + a_o^2}, \,\,\,\,\,\,
 \kappa_+ = \frac{4 m^3 (1-P^2)}{16 m^4 + a_o^2}.
\label{kpm}
\end{eqnarray}
for the surface gravity on the
inner and outer horizons.

In the last section we will also study the stability  of a second, more symmetric, space-time that
has exactly the same form as (\ref{statgmunu}) but with $r_*^2$ in place of $a_o$.

\section{A scalar field on the LBH background}
\label{scalarfield}

The wave-equation for a scalar field in a general spherically symmetric curved space-time reads
\begin{eqnarray}
\frac{1}{\sqrt{ -g }} \partial_{\mu} \left( g^{\mu \nu} \sqrt{- g} \partial_{\nu} \Phi \right) - m_{\Phi}^2 \Phi =0,
\label{SF}
\end{eqnarray}
where $\Phi \equiv \Phi(r, \theta, \phi, t)$. Inserting the metric of the
self-dual black hole we obtain the following differential equation
\begin{eqnarray}
\hspace*{0cm} && H(r) \left(2  \frac{\partial^2 \Phi }{\partial t^2}
-G(r) F'(r)  \frac{\partial \Phi }{\partial r}     \right) \nonumber \\
&& - 2 G(r) \left(   \frac{\partial^2 \Phi }{\partial  \theta^2}
+\cot  \theta   \frac{\partial \Phi }{\partial  \theta}
+\csc ^2 \theta  \frac{\partial^2 \Phi }{\partial  \phi^2}   \right) \nonumber \\
&& \hspace*{0cm} - F(r) \Big(32 m_{\Phi}^2 \csc \theta    \sqrt{\frac{G(r)}{F(r)}}   \Phi
+   H(r) G'(r)  \frac{\partial \Phi }{\partial r} \nonumber \\
&&
+2   G(r) H'(r) \frac{\partial \Phi }{\partial r}
+2 G(r) H(r) \frac{\partial^2 \Phi }{\partial r^2} \Big) = 0
 \label{SFGFH}
\end{eqnarray}
where a dash indicates a partial derivative with respect to $r$.
Making use of spherical symmetry and time-translation invariance, we write the scalar field as
\begin{eqnarray}
\Phi(r, \theta, \phi, t) := T(t) \, \varphi(r) \, Y(\theta, \phi) \quad.
\label{dec1Phi}
\end{eqnarray}
omitting the indexes $l,m$ in the spherical harmonic functions $Y_{l m}(\theta, \phi)$.
Using the standard method of separation of variables allows us to split Eq. (\ref{SFGFH})
in three equations, one depending on the $r$ coordinate, one on the
$t$ coordinate and the remaining one depending
on the angular variables $\theta, \phi$,
%
\begin{eqnarray}
&& \hspace{0cm}
\frac{\sqrt{G F}}{H} \frac{\partial}{\partial r}
\left( H \sqrt{G F} \,\,  \frac{ \partial \varphi(r)}{\partial r} \right)   \label{radial} \\
&& \hspace{-0.3cm}
= \left[ G \left(m_{\Phi}^2 + \frac{l(l+1)}{H} \right) - \omega^2 \right] \varphi(r), \nonumber\\
%
&&  \hspace{-0.3cm}
\hspace{0cm}  \left( \frac{\partial^2  }{\partial  \theta^2}
+\cot  \theta   \frac{\partial  }{\partial  \theta}
+\csc ^2 \theta  \, \frac{\partial^2  }{\partial  \phi^2}   \right) Y(\theta, \phi) 
\hspace{-0.15cm} = 
- K^2 Y(\theta, \phi), \nonumber
\end{eqnarray}
\begin{eqnarray}
 \frac{\partial^2 }{\partial t^2}  T(t) = - \omega^2 T(t),
\label{TpY}
\end{eqnarray}
where $K^2 = l(l+1)$.
To further simplify this expression we rewrite it by use of the tortoise coordinate $r^*$ implicitly defined by
\begin{eqnarray}
\frac{d r^*}{d r} := \frac{1}{\sqrt{GF}} ~.
\label{torto}
\end{eqnarray}
Integration yields the new radial tortoise coordinate
\begin{eqnarray}
&& r^* = r   - \frac{a_o^2}{r \, r_- r_+}
+ a_o^2 \frac{ \left( r_-  +  r_+ \right)}{ r_-^2
   r_+^2 } \log(r) \nonumber  \\
&&   \hspace{1cm}
- \frac{\left( a_o^2 + r_-^4\right)}{r_-^2 (r_+ - r_-)}   \log |r - r_-| \nonumber  \\
&& \hspace{1cm}
+   \frac{\left(a_o^2 + r_+^4\right) }{r_+^2
   (r_+  -  r_-)} \log |r- r_+| ~.
\label{tortoise}
\end{eqnarray}
Further introducing the new radial field $\varphi(r) := \psi(r)/\sqrt{H}$
the radial equation (\ref{radial}) simplifies to
\begin{eqnarray}\label{simply}
&& \hspace{-0.5cm} \left[\frac{\partial^2}{\partial r^{* 2}} + \omega^2 - V(r(r^*)) \right] \psi(r) = 0,  \\
&& \hspace{-0.5cm}
V(r) = G \left(m_{\Phi}^2 +\frac{ K^2 }{H} \right) 
+ \frac{1}{2} \sqrt{ \frac{G F}{H}} \left[ \frac{\partial}{\partial r} \left(  \sqrt{ \frac{G F}{H}}  \frac{ \partial H}{\partial r} \right) \right]. \nonumber
\end{eqnarray}
Inserting the metric of the self-dual black hole we finally obtain
\begin{eqnarray}
&& \hspace{-0.3cm}
V(r) = \frac{(r-r_-) (r-r_+)}{(r^4 + a_o^2)^4}
\Big[ \left(a_o^2+r^4\right)^3 m_{\Phi }^2 (r+r_*)^2 \nonumber \\
&& \hspace{-0.3cm}
+r^2 \Big(a_o^4 \left(r
   \left( \, \left(K^2-2\right) r+r_- + r_+\right)+2 K^2 r r_*+K^2 r_*^2 \right) \nonumber \\
   &&  \hspace{-0.3cm}
   +2 a_o^2 r^4
   \Big(\left(K^2+5\right) r^2+2 K^2 r r_*+K^2 r_*^2-5 r (r_-+r_+) \nonumber \\
   && \hspace{-0.3cm}
   +5 r_- r_+\Big)+r^8
   \left(K^2 (r+r_*)^2+r (r_-+r_+)-2 r_- r_+\right)\Big)  \Big]. \nonumber
   \label{VS}
   \end{eqnarray}
   For $K^2 =l(l+1) =0$
   \begin{eqnarray}
&& \hspace{-0.3cm}  V_{0}(r) =  \frac{(r-r_-) (r-r_+)}{\left(a_o^2+r^4\right)^4}
 \Big[\left(   a_o^2+r^4\right)^3 m_{\Phi }^2 (r+r_*)^2 \nonumber \\
 && 
 +r^2 \Big(a_o^4 r (-2
   r+r_-+r_+) +2 a_o^2 r^4 \big(5 r^2-5 r (r_-+r_+) \nonumber \\
   && \hspace{-0.0cm}
   +5 r_- r_+ \big)  +r^8 (r
   (r_-+r_+)-2 r_- r_+)  \Big) \Big]. \nonumber
   \end{eqnarray}
The potential $V(r)$ is zero at $r = r_+$ and $r_-$ as for the classical Reissner-Nordstr\(\ddot{\text{o}}\)m black hole.
We therefore can follow the same analysis as for this case,  approximating $V(r(r^*))$ near the horizons via
\begin{eqnarray}
&& \hspace{-0.5cm} V(r^*) \propto e^{2 \kappa_+ r^*} \,\,\,\,\,  , \,\,\, {\rm for} \,\,\,  r \rightarrow r_+ \,\,\, {\rm or} \,\,\, r^* \rightarrow - \infty \, , \nonumber \\
&&  \hspace{-0.5cm} V(r^*) \propto e^{- 2 \kappa_- r^*} \, , \,\,\, {\rm for} \,\,\,  r \rightarrow r_- \,\,\, {\rm or} \,\,\, r^* \rightarrow + \infty \, .
\label{apprV}
\end{eqnarray}

We will now focus on massless fields near the horizons. If we ignore the angular part of the solution then the field is given by
$$\psi=\int \frac{\alpha(\omega)}{r}\psi_\omega e^{-i\omega t}d\omega, $$
 where \(\alpha(\omega)\)
gives the spectrum and \(\psi_\omega\) are solutions to Eq. (\ref{simply}). \(\psi_\omega e^{-i\omega t}\) will in general consist of two linearly independent solutions corresponding to right-moving (outgoing) and left-moving (ingoing) waves traveling along surfaces of constant null coordinates \(u=r^*-t\) and \(v=r^*+t\) respectively. Thus, the total field \(\psi\) can be decomposed into two functions, one of \(u\) and one of \(v\), which describe its right-moving and left-moving modes. We represent this with the equation
\begin{align} \label{total}
    \psi = \frac{1}{r}\left[g^{(-)}(u)+g^{(+)}(v)\right].
\end{align}

We will need to derive the form of \(g^{(-)}\) and \(g^{(+)}\) eventually. Assuming that they are given, however, we can compute the energy density \(\rho\) of the field as measured by a freely falling observer near the horizon with four-velocity \(U^\alpha\); we have
$$\rho=\psi_{, \alpha} \psi_{, \beta} U^\alpha U^\beta+\frac{1}{2}\psi_{, \alpha} \psi^{*,\alpha}. $$
However, since \(u,v=\text{const}\) are null surfaces, the form of \(\psi\) near the horizons implies that this will be dominated by the \(|\psi_{,\alpha}U^\alpha|^2\) term.

The four-velocity of a timelike, radial geodesic is  
\begin{eqnarray}
&&  \hspace{-0cm}  U^t =\frac{E}{G}=\frac{E(r^4+a_o^2)}{(r-r_+)(r-r_-)(r+r_*)^2} \nonumber \\
 &&  \hspace{-0cm} U^r =-\left[\frac{F}{G}\left(E^2-G\right)\right]^{1/2}   \\
&&    \hspace{-0.0cm} =\frac{-r^2}{(r+r_*)^2}\left[E^2-\frac{(r-r_+)(r-r_-)(r+r_*)^2}{r^4+a_o^2}\right]^{1/2}.
\nonumber
\end{eqnarray}
 where we define \(U^r\) to be negative;  this will always be the case between the horizons \(r_- < r < r_+\) (since \(r\) necessarily decreases for any observer in this region). \(E\) is a constant of the motion, the sign of which gives the direction of travel between the horizons; \(E>0\) corresponds to a left-moving observer and \(E<0\) to a right-moving one (This can be seen because between the horizons \(G\) will be negative and the \(t\)-coordinate runs to the right).

Our goal is to compute the energy density \(\rho \propto |U^\alpha g^{(\pm)}_{,\alpha}|^2\). It is easily seen that \(g^{(\pm)}_{,t}=\pm g^{(\pm)'}\) and \(g^{(\pm)}_{,r}=\frac{1}{\sqrt{GF}}g^{(\pm)'}\), where \(g^{(\pm)'}\) is the derivative of \(g^{(\pm)}\) with respect to \(v\) or \(u\) depending on the sign. With this we obtain
\begin{eqnarray}
   && \hspace{-0.5cm}
   U^\alpha g^{(\pm)}_{,\alpha} = \frac{g^{(\pm)'}}{G}\left(\pm E - |E^2-G|^{1/2}\right) \nonumber \\
    &&  \hspace{-0.5cm}
    =\frac{(r^4+a_o^2)g^{(\pm)'}}{(r-r_+)(r-r_-)(r+r_*)^2} \times \nonumber \\
     &&   \hspace{-0.5cm}
     \left(\pm E - \left|E^2-\frac{(r-r_+)(r-r_-)(r+r_*)^2}{r^4+a_o^2}\right|^{1/2}\right).
\end{eqnarray}

Let us examine this result in the limit \(r\rightarrow r_-\). For a left-moving observer (\(E>0\)) we see that \(U^\alpha g^{(+)}_{,\alpha}\) remains finite but \(U^\alpha g^{(-)}_{,\alpha}\) diverges, unless of course \(g^{(-)'}\) can compensate for the divergence. Conversely, for a right-moving observer (\(E<0\)) \(U^\alpha g^{(+)}_{,\alpha}\) diverges while \(U^\alpha g^{(-)}_{,\alpha}\) remains finite.

Near the inner horizon, \(r\simeq r_-\), the tortoise coordinate is dominated by
\begin{eqnarray}
   \hspace{0.1cm}
    r^*\simeq \frac{r^4+a_o^2}{r_-^2(r_--r_+)}\log |r-r_-| = -(2 \kappa_-)^{-1}\log |r-r_-| .  \nonumber
\end{eqnarray}
From this is can be shown that for a right-moving observer \(\frac{dr^*}{dt}=\frac{dr^*}{dr}\frac{U^r}{U^t} \simeq 1\), from which we obtain
\begin{eqnarray}
  && \hspace{-0.5cm} -v=-t-r^*\simeq -2r^*+\text{const}
   \\
 && \hspace{-0.5cm}  \simeq \kappa_-^{-1}\log|r-r_-|+\text{const} 
    \implies 
 \,\,\,\,    (r-r_-)^{-1} \propto e^{\kappa_- v} .  \nonumber
\end{eqnarray}

This gives us the form of the divergence in \(U^\alpha g^{(+)}_{,\alpha}\) expressed in null coordinates (recall that as \(r\rightarrow r_-\), \(v\rightarrow \infty\) for a right-moving observer and \(u \rightarrow \infty\) for a left-moving one). That is,
\begin{align}
    U^\alpha g^{(+)}_{,\alpha} \propto g^{(+)'} e^{\kappa_- v} \;\;\; \text{for} \;\; r\simeq r_-
    \;\; \text{and} \;\; E<0 .
\end{align}
Thus, if the loop black hole is to remain stable on \(r_-\) then \(g^{(+)'}\) must decay at least as fast as \(e^{-\kappa_- v}\) in order to placate the divergence as \(v\rightarrow \infty\). A similar analysis shows that \(g^{(-)'}\) must decay at least as fast as \(e^{-\kappa_- u}\) in order to stop the divergence of \(U^\alpha g^{(-)}_{,\alpha}\) as \(u\rightarrow \infty\) for observers with \(E>0\). Our goal now is now to compute these quantities to determine stability.

\section{Near horizon solution}
\label{results}

\begin{figure}
 \centering
  \includegraphics[width=7.5cm]{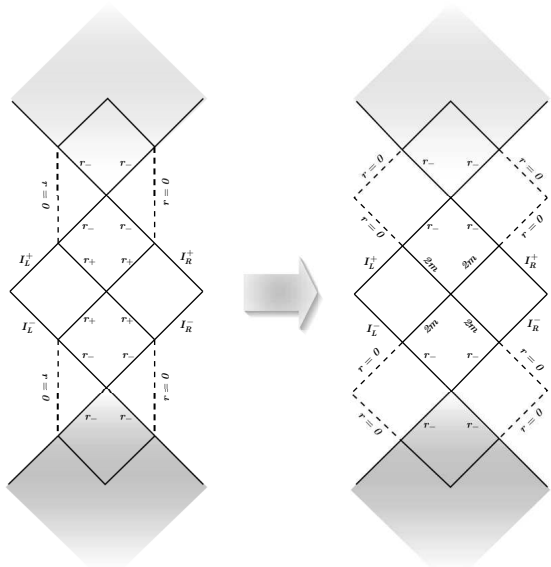}
  \caption{Penrose diagram for the loop black hole on the right and its Reissner-Nordstr\(\ddot{\text{o}}\)m analog on the left. }
\label{PenroseD}
  \end{figure}

In order to compute \(g^{(\pm)'}\) we reproduce a calculation used in \cite{RNstability} to determine the inner horizon stability of the Reissner-Nordstr\(\ddot{\text{o}}\)m black hole. The applicability of this same calculation to the present case is due to the similarities between the Reissner-Nordstr\(\ddot{\text{o}}\)m spacetime and that of the loop black hole; both spacetimes are displayed in Fig. (\ref{PenroseD}).

Since we are only interested in the field near the horizons, where the potential is exceedingly small, we can decompose the total solution to Eq. (\ref{simply}), which we will call \(\Psi_\omega\), into the zero-potential solution \(\psi_\omega\) plus an infinitesimal perturbation produced by the small potential \(\epsilon_\omega\): \(\Psi_\omega=\psi_\omega+\epsilon_\omega\). We set the initial conditions to consist only of ingoing waves and we choose the time dependence to be \(e^{-i\omega t}\); this requires that \(\psi_\omega=e^{-i\omega r^*}\).

\begin{figure}
 \centering
  \includegraphics[width=7.5cm]{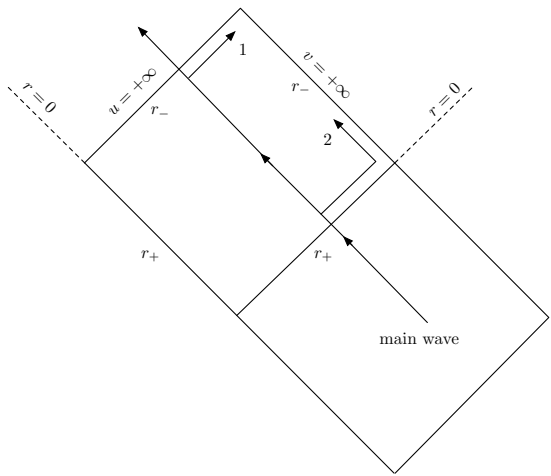}
    \caption{Displaying the two scatterings off of the main wave which could potentially cause instability at the Cauchy horizon. For scatterings 1 and 2 we must check the forms of \(g^{(-)'}\) and \(g^{(+)'}\) respectively.}
\label{scat}
  \end{figure}

An ingoing (left-moving) wave will scatter off of the small potential near the horizons, and these scatterings are represented by \(\epsilon_\omega\). There are two scatterings that are of potential interest with regard to stability at the inner horizon, as displayed in Fig. (\ref{scat}). The first of these consists of right-moving waves traveling along the left branch of \(r_-\); these would have scattered off of the main wave as it neared \(r_-\) and are labeled with a ``1" in Fig. (\ref{scat}). For these waves we must check the form of \(g^{(-)'}\) to test for stability. The second scattering of interest consists of left-moving waves that travel along the right branch of \(r_-\). These waves can form by the following process. Consider a scattering produced just after the main wave has entered the outer horizon; it would be right-moving and traveling along \(r_+\). As this wave approaches the intersection of \(r_+\) and \(r_-\) in the Penrose diagram it enters a region of strong potential. Seeing as the wave is assumed to be very small, a strong potential would be expected to scatter this wave again with effectively 100\% efficiency; in this case the entire wave is scattered such that it is now a left-moving wave traveling along the right branch of \(r_-\) as labeled by a ``2" in Fig. (\ref{scat}). For these waves we must check the form of \(g^{(+)'}\) to test for stability.

In order to solve for \(\epsilon_\omega\) we use a Green's function \(G_\omega(r^*,y^*)\),
\begin{align}
    \left(\frac{\partial^2}{\partial r^{*2}}+\omega^2\right)G_\omega(r^*,y^*)=\delta (r^*-y^*) .
\end{align}
 Having chosen an \(e^{-i\omega t}\) time dependence, we opt for the solution
\begin{align}
    G_\omega(r^*,y^*)=
    \begin{cases}
        \frac{1}{2i\omega}e^{i\omega(r^*-y^*)} & \text{if} \;\; r^* > y^* ,  \\
        \frac{1}{2i\omega}e^{-i\omega(r^*-y^*)} & \text{if} \;\; r^* < y^* .
    \end{cases}
\end{align}

Now, since \(\Psi_\omega=\psi_\omega+\epsilon_\omega\) solves Eq. \ref{simply} while \(\psi_\omega\) solves the same equation with the potential set to zero, we find that \(\epsilon_\omega\) approximately solves
\begin{align}
    \left(\frac{\partial^2}{\partial r^{*2}}+\omega^2\right)\epsilon_\omega(r^*)=V(r^*)\psi_\omega(r^*) ,
\end{align}
the solution of which is
\begin{align}
    \epsilon_\omega(r^*)=\int_{-\infty}^\infty G_\omega(r^*,y^*)V(y^*)\psi_\omega(y^*)dy^* .
\end{align}

Here we are interested in the scattering produced by the outer horizon potential \(V(y^*)=V_0e^{2\kappa_+ y^*}\), \(y^* \rightarrow -\infty\). For purposes of computational ease let us set \(V(y^*)=0\) for \(y^* \geq 0\). We evaluate this integral for \(r^*>0\) since this will be the case when the wave approaches \(r_-\), which is what we are interested in. Identifying \(\psi_\omega=e^{-i\omega r^*}\) we obtain trivially
\begin{align}
    \epsilon_\omega (r^*)=\frac{V_0 e^{i\omega r^*}}{4i\omega(\kappa_+-i\omega)} .
\end{align}
Including time dependence we obtain the right-moving wave
\begin{align}
    e^{-i\omega t}\epsilon_\omega (r^*)=\frac{V_0 e^{i\omega u}}{4i\omega(\kappa_+-i\omega)} .
\end{align}

We now need to evaluate the total wave consisting of all modes. Let us impose a \(\delta\)-function pulse for the form of the primary wave \(\psi=\int \frac{\alpha(\omega)}{r}\psi_\omega e^{-i\omega t}d\omega=\delta(v)/r\); this implies \(\alpha(\omega)=\frac{1}{2\pi}\) and thus gives the total scattered wave
\begin{align}
   \hspace{-0.2cm}  \epsilon=\frac{1}{2\pi r}\int_{-\infty}^\infty e^{-i\omega t} \epsilon_\omega d\omega
    =\frac{1}{2\pi r}\int_{-\infty}^\infty \frac{V_0 e^{i\omega u}}{4i\omega(\kappa_+-i\omega)} d\omega.
\end{align}

As explained above, when this wave approaches the intersection of \(r_+\) and \(r_-\) in the Penrose diagram it will be entering a region of high potential, which can be expected to scatter the small wave with near 100\% efficiency. We assume this condition here, and so the wave after this second scattering will be of the same form as that just derived but traveling leftwards instead of rightwards. Recalling Eq. \ref{total}, we thus have the form of \(g^{(+)}(v)\) for this wave
\begin{align}
    g^{(+)}(v)=\frac{1}{2\pi}\int_{-\infty}^\infty \frac{V_0 e^{-i\omega v}}{4i\omega(\kappa_+-i\omega)} d\omega .
\end{align}
where the exponential is negative because we are still using an \(e^{-i\omega t}\) time dependence.

In order to test for stability, however, we require the derivative of this:
\begin{align}
    g^{(+)'}(v)= - \frac{1}{2\pi}\int_{-\infty}^\infty \frac{V_0 e^{-i\omega v}}{4(\kappa_+-i\omega)} d\omega .
\end{align}
This is evaluated via a simple contour integration and gives the primary result of this paper
\begin{align}\label{gvsol}
    g^{(+)'}(v)= - \frac{V_0}{4}e^{-\kappa_+ v} .
\end{align}

Recall that the stability of the inner horizon is contingent on this quantity decaying at least as fast as \(e^{-\kappa_- v}\), and so we see that it simply comes down to comparing the two surface gravities \(\kappa_+\) and \(\kappa_-\). For the Reissner-Nordstr\(\ddot{\text{o}}\)m black hole this result was used to show that the inner horizon is unstable \cite{RNstability}, since in that case one always has \(\kappa_- > \kappa_+\), and so \(g^{(+)'}(v)\) does not decay fast enough to suppress the energy density divergence. This is not necessarily the case for the loop black hole, however.

Recall that there were two possible divergences that could occur at \(r_-\), the other one being generated by fields near the left branch of \(r_-\) and with stability contingent on \(g^{(-)'}(u)\) decaying at least as fast as \(e^{-\kappa_- u}\). In this case a similar analysis can be performed which shows that it decays \emph{exacly} as fast this, thus maintaining stability in this respect \cite{RNstability}. We therefore consider only Eq. (\ref{gvsol}) in the sequel.

The two surface gravities are given by
\begin{align}
    \kappa_+&=\frac{4m^3(1-P^2)}{16m^4+a_o^2}=\frac{r_+^2(r_+-r_-)}{2(r_+^4+a_o^2)}, \\
    \kappa_-&=\frac{4m^3P^4(1-P^2)}{16m^4P^8+a_o^2}=\frac{r_-^2(r_+-r_-)}{2(r_-^4+a_o^2)},
\end{align}
where \(r_+=2m\), \(r_-=2mP^2\) and the ADM mass of the black hole is \(M=m(1+P)^2\).

Unlike the Reissner-Nordstr\(\ddot{\text{o}}\)m black hole it is entirely possible here to have \(\kappa_- < \kappa_+\), which we have shown to result in a stable inner horizon. To see this recall that \(P\) is expected to be a very small number, and in the limit \(P\rightarrow 0\) we have \(\kappa_- \rightarrow 0\) while \(\kappa_+\) remains non-zero. The limiting case for stability is when the surface gravities are equal
\(\kappa_+=\kappa_-\); under this condition we refer to the loop black hole as being symmetric since both an observer and a dual observer will see the exact same metric. \(\kappa_+=\kappa_-\)
(see next section).
%
%
We see that stability of the inner horizon will be retained (\(\kappa_-\leq \kappa_+\)) only as long as
\begin{align}
    M\leq \frac{\sqrt{a_o}(1+P)^2}{2P} .
    \end{align}
That is, with a small enough (but not zero) polymeric parameter $P$ the loop black hole will be stable.

With this inequality we can give a heuristic upper bound for \(P\) by setting \(a_o\) to the Planck area and \(M\) to the estimated mass of the universe \(M\sim 10^{53}\) kg \cite{universemass}. Requiring stability even in this extreme mass case gives a bound of \(P \lesssim 10^{-61}\). Assuming that the Immirzi parameter is on the order of unity \(\gamma \sim 1\) this further gives a bound on the polymeric parameter \(\delta \lesssim 10^{-30}\).  Note that setting these parameters to these values renders all  LBHs in our universe stable.  However it does not render all possible LBHs stable.
In the next section we consider a subclass of LBHs that are fully stable under the perturbations we consider.

\section{The symmetric black hole} 

In this section we consider a different metric, obtained by fixing in a different way
the integration constant $B$ appearing in the general solution, which reads  
\begin{eqnarray}
&& \hspace{-0.5cm} ds^2 = -\frac{ (r - r_+) (r-r_-) (r+ r_*)^2 }{ r^4 + {B}^2}dt^2
\nonumber \\
&& \hspace{0.2cm} +\frac{dr^2}{\frac{(r-r_+)(r-r_-)r^4}{(r+ r_*)^2 (r^4 +  {B}^2)}}
+ \Big(\frac{ B^2}{ r^2} + r^2\Big) d\Omega^{(2)}.
\label{metricabella}
\end{eqnarray}
The metric considered in the previous section is obtained fixing $B$ (referred to as  
the bounce parameter) using
 the minimum area of the full theory (LQG); for more details see \cite{Modesto:2008im}.
 Here we instead fix this parameter in such a way as to utilize the dual nature of the semiclassical metric.
We recall that the metric presents two event horizons
in $r_+ = 2m $ and $r_-=2 m P^2$ and two free parameters:
the polymeric function $P$,
which is a function of the product $\gamma^2  \delta^2$, and the
free bounce parameter $B$, which has dimensions of (length)$^2$. 
In the limit $P \rightarrow 0$ and $B \rightarrow 0$ the metric
reduces to the Schwarzschild solution.

Going back to the property of self-dualty, 
the metric (\ref{metricabella}) is invariant (in form) under the transformation
\begin{eqnarray}
r \rightarrow R =\frac{B}{r}
\end{eqnarray}
and the dual metric is 
\begin{eqnarray}
&& \hspace{-0.5cm} ds^2 = -\frac{ (R - R_+) (R-R_-) (R+ R_*)^2 }{R^4 + B^2}dt^2
\nonumber \\
&& \hspace{0.2cm} +\frac{dR^2}{\frac{(R-R_+)(R-R_-)r^4}{(R+ R_*)^2 (R^4 +  B^2)}}
+ \Big(\frac{ B^2}{ R^2} + R^2\Big) d\Omega^{(2)},
\label{metricabellad}
\end{eqnarray}
if we define $R_+ = B/2 m P^2$, $R_- = B/2 m$
and $R_* = B/2 m P$ and we redefine the time coordinate
to $t \rightarrow t \, r_*^2/ B$.

Now we fix the bounce parameter $B$ such that the duality is upgraded to a \emph{symmetry} of the metric.
The metric is invariant under the symmetry
$r \rightarrow  R = B/r$ iff $B = r_*^2$, namely
\begin{eqnarray}
g_{\mu\nu}(r) \rightarrow g'_{\mu \nu}(R) = g_{\mu \nu}( R )  \,\,\, \forall \, R.
\end{eqnarray}
 In this case it is not necessary to redefine
the time coordinate. Furthermore, the dual observer sees exactly the same mass $m$ because
$R_+ = 2m = r_+$, $R_- = 2 m P^2 = r_-$ and $R_* = 2 m P = r_*$.
The final form of the metric is
\begin{eqnarray}
&& \hspace{-0.5cm} ds^2 = -\frac{ (r - r_+) (r-r_-) (r+ r_*)^2 }{ r^4 + {r_*}^4}dt^2
\nonumber \\
&& \hspace{0.2cm} +\frac{dr^2}{\frac{(r-r_+)(r-r_-)r^4}{(r+ r_*)^2 (r^4 +  {r_*}^4)}}
+ \Big(\frac{ {r_*}^4}{ r^2} + r^2\Big) d\Omega^{(2)}.
\label{metricabelladd}
\end{eqnarray}
On the other hand we can expand the component $g^{rr}(r)$ of the metric and obtain the
ADM mass from the coefficient of the term that is  first order in $1/r$. 
The result is $m_{ADM} = m (1 + P)^2$.
In other words the ADM mass and the dual ADM mass are equal.
For this solution the surface gravity on the event horizon
is equal to the surface gravity on the Cauchy horizon
\begin{eqnarray}
\kappa_+ = \kappa_- =  \frac{(1-P^2)}{4 m (1 +P^4)}
\end{eqnarray}
and based on our previous analysis the Cauchy horizon is stable $\forall \, P$.
The interesting result in this section is that the new symmetry between
large and short distances protects the inner Cauchy horizon from collapsing
to a curvature singularity. We have the combination of two effects:
the presence of the polymeric parameter $P$ which regularizes the metric
and the new symmetry. It seems both are necessary to have a stable
Cauchy horizon for any black hole mass.


\section{Conclusions}

We have studied the stability of the loop black hole by considering perturbative scatterings off of an ingoing field pulse. Applying the same analysis performed in \cite{RNstability} it was found that the energy density of these perturbations diverges to linear order near the Cauchy horizon if the surface gravities satisfy \(\kappa_- > \kappa_+\). It should be noted that since the perturbations are assumed very small such a divergence indicates a breakdown in our approximation. As such, while this divergence is highly indicative of instability it is not conclusive. Using the metric Eq. (\ref{g}) we discovered that Cauchy horizon stability (in the sense just described) is contingent on the mass of the black hole being less than a specified value, the magnitude of which is determined by the constants of the underlying theory. We also found that stability can be achieved independent of the black hole mass by making a different choice of the constant \(B\) present in the derivation of the metric.

The analysis presented here can be easily performed for other quantum gravity inspired black hole solutions \cite{NCBH} since they all seem to have a curiously similar spacetime structure to that of the Reissner-Nordstr\(\ddot{\text{o}}\)m black hole. In the end it comes down to simply comparing the surface gravities of the horizons: if \(\kappa_- > \kappa_+\) then the Cauchy horizon stability becomes questionable. For example, a similar analysis has been performed for a black hole metric inspired by noncommutative geometry \cite{bh3,bh4}; this work will be available in a future publication. In this case it was found that the Cauchy horizon is always unstable, even when an ultraviolet cutoff is added to the field.

A fundamental result in the theory of Cauchy horizon instability is the phenomenon of mass inflation. This was discovered by Poisson and Israel \cite{poisson}; it is a process in which the mass function of the Reissner-Nordstr\(\ddot{\text{o}}\)m black hole diverges at the Cauchy horizon and is considered to be a much stronger and conclusive argument towards instability than what we have presented here. Applying the more rigorous analysis of mass inflation to the loop black hole represents the next step in this line of research, and it is the path that the authors now plan to take.

Another approximation made in this analysis, and one present in the derivation of mass inflation as well, is that we have not bothered to quantize the matter fields propagating in the black hole. It is largely unclear what changes such a quantization would make on the result of Cauchy horizon instability \cite{poisson2}.

Finally, we wish to ponder whether or not the discovery of a stable Cauchy horizon should be an encouraging one or not. The primary goal of developing quantum gravity black hole solutions seems to be to placate the singularities present in their classical counterparts. As such, the result of a stable Cauchy horizon would be seen as a success among quantum gravity theorists. It must be remembered, however that relativists breathed a great sigh of relief upon the discovery that the Cauchy horizon was indeed (classically) unstable \cite{poisson}. This is because without such a singularity the Cauchy horizon becomes traversable and we are left with the result that our theory is no longer deterministic. In classical general relativity this is seen as a \emph{big} problem, as well it should. Of course when one includes quantum effects it may not be surprising or even disturbing to find that the theory becomes nondeterministic, but this perspective could likely be debated given the context under which we lose determinism here. We conclude that further thought and insight is needed to solve this quandary.

\section*{Acknowledgements}

L M thanks Alberto Montina for the assistance given in a preliminary calculation. 
Research at
Perimeter Institute is supported by the Government of Canada through Industry Canada
and by the Province of Ontario through the Ministry of Research \& Innovation.  This work
was supported in part by the Natural Sciences \& Engineering Research Council of Canada.

\end{document}